# From M100 to NGC 7331: A Multiband Study and Physical Parameters in Nearby Galaxies

Bertua Marasca, J. C.[1]; Dávila Gutiérrez, H.[1,2]; Huaman Ticona, V.[1]; Mosquera Hadatty, J. I.[1]; Ramírez Valadez, A. A.[1]

[1] Master in Astrophysics, Universidad Internacional de La Rioja (UNIR), 26006 Logroño, Spain
[2] SKYCR.ORG, San José, Costa Rica

**Abstract**

We study five nearby galaxies (M100/NGC 4321, NGC 1300, M 74, M 60, and NGC 7331) by combining multiband imaging (optical, UV, NIR, and X-rays) with simple photometric measurements to show how each spectral window traces different physical components: UV/blue emphasizes recent star formation (Gil de Paz et al. 2007), NIR outlines the old stellar mass and internal structure (Elmegreen et al. 2004; Ryder et al. 1998), and X-rays reveal compact sources and hot gas (Palumbo et al. 1981; Kaaret 2001; ESA 2005). For M100 we present a multiband mosaic with X-ray contours including the SN 1979C region (Kaaret 2001). We compare isophotal morphology in NGC 1300 (barred) and M 74 (unbarred), quantify ellipticity and equivalent radius in M 60 (Tonry & Schneider 1988), and illustrate a Tully–Fisher distance estimate for NGC 7331, discussing assumptions (inclination, internal corrections) and consistency with classical scales (Rubin et al. 1965; HST Key Project 2000). Our results underscore that a "layered" view reduces interpretative biases and that straightforward integrations of public data with open-source software yield robust introductory-level physical diagnostics (Knapen et al. 1999).

**Context.** Each spectral window traces different physical components: UV/blue emphasizes recent star formation; the near-infrared (NIR) follows the old stellar mass and internal structure; and X-rays reveal compact energetic sources and hot gas (Gil de Paz et al. 2007; Elmegreen et al. 2004; Ryder et al. 1998; Palumbo et al. 1981; Kaaret 2001; ESA 2005).
**Aims.** We provide an instructive, reproducible comparison across five nearby galaxies—M100/NGC 4321, NGC 1300, M 74, M 60, and NGC 7331—to show how multiband imaging constrains morphology and basic physical parameters, including distances.
**Methods.** We combined public UV, optical, NIR, and X-ray images and performed simple photometric and geometric measurements. For M100 we assembled a multiband mosaic with X-ray contours that include the SN 1979C region (Kaaret 2001). We compared isophotal morphology in the barred NGC 1300 and the unbarred M 74, and quantified ellipticity and equivalent radius in M 60 (Tonry & Schneider 1988). For NGC 7331 we illustrated a Tully–Fisher distance estimate, discussing assumptions (inclination, internal corrections) and consistency with classical scales (Rubin et al. 1965; HST Key Project 2000).
**Results.** A "layered" view mitigates interpretative biases: UV/blue highlights star-forming regions (Gil de Paz et al. 2007), NIR delineates the old stellar mass and internal structure (Elmegreen et al. 2004; Ryder et al. 1998), and X-ray maps expose compact sources and hot gas (Palumbo et al. 1981; Kaaret 2001; ESA 2005). Straightforward integrations of public data with open-source software provide robust introductory-level physical diagnostics (Knapen et al. 1999), and the Tully–Fisher distance for NGC 7331 is consistent with classical determinations (Rubin et al. 1965; HST Key Project 2000).

**Key words.** galaxies: spiral — galaxies: structure — galaxies: photometry — galaxies: distances and redshifts — X-rays: galaxies — techniques: photometric.

## Introduction

Understanding a galaxy requires a layered view. The ultraviolet (UV) traces the most recent star formation (Gil de Paz et al. 2007); the optical outlines the global morphology; the near-infrared (NIR), being less affected by dust, approximates the distribution of the old stellar mass and the internal geometry (Elmegreen et al. 2004; Ryder et al. 1998); and X-rays reveal compact binaries, supernova remnants, and hot gas (Palumbo et al. 1981; Kaaret 2001). Our sample—M100/NGC 4321, NGC 1300, M 74/NGC 628, M 60/NGC 4649, and NGC 7331—covers a grand-design spiral, a barred versus an unbarred spiral, a luminous elliptical, and a massive, well-studied spiral, with morphological types taken from standard catalogues (de Vaucouleurs 1976). For M100, multiband studies of the central environment are available (Knapen et al. 1999; Tonry & Schneider 1988), as well as X-ray analyses (Palumbo et al. 1981; ESA 2005; Kaaret 2001). For NGC 7331, classical kinematics (Rubin et al. 1965) and Cepheid/Key-Project distance determinations are available (HST Key Project 2000).

In this work we present an instructive, reproducible comparison across five nearby galaxies of morphological interest: M100 (NGC 4321), NGC 1300, M 74 (NGC 628), M 60 (NGC 4649), and NGC 7331. The sample was selected to encompass: (i) a grand-design spiral rich in young tracers (M100); (ii) a strongly barred spiral (NGC 1300) contrasted with an unbarred grand-design spiral (M 74); (iii) a bright elliptical for isophotal and ellipticity analysis (M 60); and (iv) a massive, well-studied spiral suitable for illustrating a Tully–Fisher distance estimate (NGC 7331).

Methodologically, we combine public UV, optical, NIR, and X-ray images obtained from open archives and processed/registered with open-source software. On these products we perform simple geometric measurements (isophotes; major and minor axes; ellipticity, $\varepsilon = 1 - \frac{b}{a}$; and equivalent radius, $R_{eq} = \sqrt{ab}$ ), and construct cross-band morphological comparisons to distinguish structures that trace mass from those that are merely luminous tracers of young populations. Where appropriate, we use the isophotal radius $R_{25}$ (defined at $\mu B = 25 \, mag \, arcsec^{-2}$) as an operational scale.

Our approach has three complementary goals. First, to show how multiband inspection modifies—and often corrects—morphological interpretations based solely on the visible: prominent blue arms do not necessarily coincide with the NIR stellar-mass distribution. Second, to quantify, via isophotes, the presence or absence of a bar—a key driver of angular-momentum redistribution and central fueling—by comparing NGC 1300 with M 74. Third, to illustrate a field-galaxy distance estimate with the Tully–Fisher relation in NGC 7331, making explicit the main sources of uncertainty (inclination, internal extinction, and band-dependent calibration).

The paper is organized as follows. In Sect. 2 we describe the data sources, preprocessing (registration, common scale, and standard N–E orientation), and metrics. In Sect. 3 we present the multiband view of M100 (including X-ray contours), contrast the isophotal morphology of NGC 1300 and M 74, measure the ellipticity and equivalent radius of M 60, and derive an example Tully–Fisher distance for NGC 7331. Sect. 4 discusses biases and limitations (extinction, age–metallicity degeneracies, and cross-survey resolution mismatches), and Sect. 5 summarizes the main conclusions and the usefulness of these techniques, based on public data and open tools, for teaching and research initiation.

## 2. Data sources

### 2.1. Datasets used in this work

In the optical (DSS2) the spiral arms with H II regions are clearly delineated; in the UV (GALEX) bursts of recent star formation stand out; in the NIR (2MASS) the light from old stellar populations dominates, tracing the bulge and possible oval/bar structure; and in X-rays (RASS/XMM/Chandra) we detect point sources (e.g. SN 1979C) together with diffuse hot gas. The UV–NIR comparison shows that the eye-catching blue spiral pattern does not necessarily trace the stellar-mass distribution.

**Optical**
– **DSS2** (photographic B, R, I; ~0.35–0.9 μm). Cutouts around each target, useful for global morphology and wide field.
– **SDSS** (where available; u,g,r,i,zu,g,r,i,zu,g,r,i,z; 0.35–0.9 μm). Improved photometry and astrometry relative to DSS2.

**Ultraviolet**
– **GALEX** (FUV 1344–1786 Å; NUV 1771–2831 Å). Traces recent star formation and OB associations.

**Near-infrared**
– **2MASS** (J 1.25 μm, H 1.65 μm, KsK_sKs 2.16 μm). Minimizes dust bias and approximates the stellar-mass distribution.

**X-rays**
– **ROSAT** All-Sky Survey (RASS; 0.1–2.4 keV): wide-field coverage for diffuse emission and bright sources.
– **XMM-Newton** EPIC (0.2–12 keV) and **Chandra** ACIS (~0.5–7 keV), when pointed observations are available.

Metadata and bibliography

- **SIMBAD:** identifications, morphological types, positions, and key references.
- **NASA/ADS:** articles and calibrations (e.g. band–dependent Tully–Fisher relations).

*Practical notes.* All images were retrieved from public archives and visualised in Aladin (CDS). When multiple products exist per band, we selected the dataset with the best resolution–coverage compromise to define a common field of view for each galaxy.

## 2.2. Preprocessing: registration, common scale, scale bar, N–E orientation, and photometry/isophotes

### (a) Field definition and data retrieval

1. We adopted a common angular field per galaxy (typically 15′–25′, depending on the apparent size $D_{25}$).
2. We downloaded DSS2/SDSS, GALEX, and 2MASS cutouts with the same centre and size. For X-rays:
– RASS: count-rate/surface-brightness maps for the same field.
– XMM/Chandra (when available): broad-band combined images.

### (b) Astrometric registration and common grid

3. A **reference image** was chosen (by default, 2MASS KsK_sKs or SDSS iii). All other images were reprojected to its WCS and pixel scale (1–2″ px$^{-1}$, set by the worst resolution in the set) using Lanczos-3 interpolation.
4. The alignment was verified against field stars, yielding a typical astrometric RMS ≲1″.

### (c) Visual homogenisation and PSF

5. For comparative figures we did not enforce strict PSF matching (this is stated in the captions).
6. For geometric measurements (isophotes; aaa, bbb, ellipticity ε\varepsilonε), we worked on the NIR image (2MASS KsK_sKs or HHH), which is less affected by dust, and transferred the same ellipses to the other bands to compare profiles with identical geometry.
*Optional (advanced):* match the PSF to the worst FWHM—typically GALEX—via Gaussian convolution when extracting fine radial profiles.

### (d) Scale bar, orientation, and annotations
7. All panels are oriented with **N up** and **E to the left** (WCS rotation).
8. We add an **angular scale bar** (e.g. 2′). When an adopted distance DDD is available, we also display a **physical scale** (kpc) computed as

$$1'' \approx 4.848 pc \times \left(\frac{D}{Mpc}\right)$$

…and we quote the adopted distance **D** and its reference in the caption.

### (e) X-ray contours
9. We smooth the X-ray map with a Gaussian filter (with σ\sigmaσ matched to the instrument resolution) and generate isocount contours at levels ($\bar{x} + n\sigma$), overlaid on the optical/NIR image.

### (f) Isophotal photometry and geometric metrics
10. We extract elliptical isophotes from the NIR image to obtain the major axis *a*, minor axis *b*, and the position angle (PA) of the major axis; when appropriate, we also report the isophotal radius $R_{25}$ (defined at $\mu B = 25\ mag\ arcsec^{-2}$), provided a B-band calibration is available.

11. We compute:

$$\epsilon \equiv 1 - \frac{b}{a},\ R_{eq} \equiv \sqrt{ab}$$

12. For radial profiles we compare relative **i**ntensities band by band within the same ellipses. If an image is not calibrated in surface brightness, we state explicitly that values are relative (normalized ADU).

### (g) Quality control and uncertainties
13. We repeat the measurement on adjacent isophotes to estimate the dispersion in *a* and *b*, and propagate it to σ(ε) and σ($R_{eq}$).
14. Main sources of error are the choice of isophote (S/N and sky subtraction), inter-band PSF differences, masking of foreground stars and satellites, and—if }$R_{25}$ is used—the B-band photometric calibration.

## 3. Results

- Public images in the optical**,** NIR**,** blue (B), and X-ray were visualised with Aladin.
- For each band, the best available product was selected and compared over the same field of view.
- All figures include a scale bar, standard N–E orientation, and data credits.

*Fig. 1.—* *Multiband mosaic of M100 (optical/NIR/UV) with X-ray contours.*

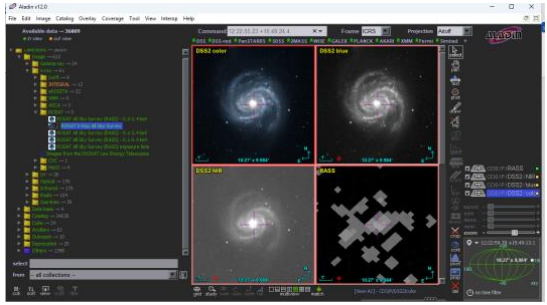
Credits. Aladin v12.0; authors' processing.

### 3.1. Optical (blue/visible band)

- Spiral arms are clearly delineated, with prominent H II regions and young clusters.
- Dust can obscure inner zones, producing irregular patchy structures.

*Fig. 2.— M100 in the optical.*

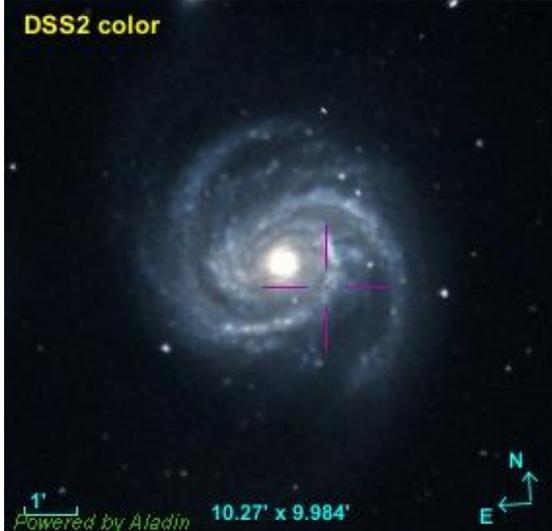
**Credits.** DSS2 + Aladin.

In the blue band the spiral arms and H II regions with young clusters are clearly visible. O–B stars emit strongly in the UV and blue and fade quickly as they age; therefore the blue light traces **recent star formation**. Dust can obscure inner zones, producing irregular patchy structures.

### 3.2. Optical (blue)

- Dominated by light from old stellar populations (bulge and bar/oval).
- Less sensitive to dust, therefore it traces the stellar mass and the internal geometry more faithfully.

*Fig. 3.— M100 in blue (DSS2).*

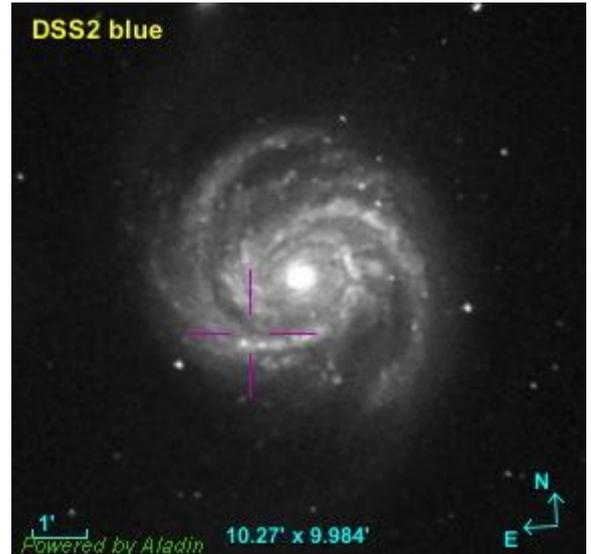
Credits. 2MASS + Aladin (CDS).

### 3.3. Near-infrared (NIR)

- Light is dominated by old stellar populations, outlining the bulge and any bar/oval.
- Less affected by dust; a better tracer of the stellar-mass distribution and internal geometry.

*Fig. 4.— M100 in the NIR (2MASS J/H/Ks).*

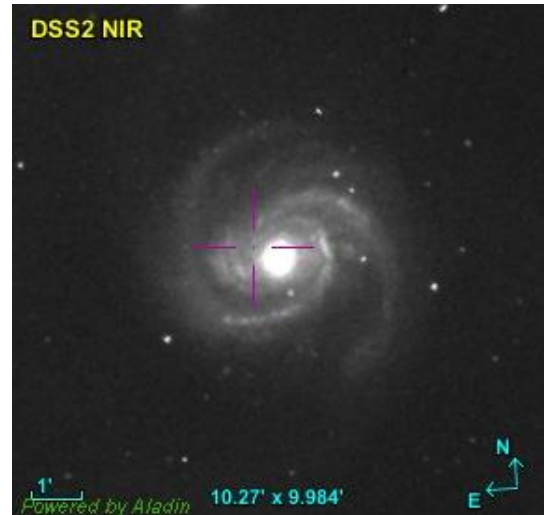
Credits. 2MASS + Aladin (CDS).

In the near-infrared (NIR) the light is dominated by old stellar populations (red giants and evolved stars), especially in the bulge and inner structures (bar/oval). The NIR is less sensitive to dust, thus tracing the stellar-mass distribution and the geometry of the nucleus and disc more faithfully. Locally, heated dust and H2\mathrm{H}_2H2 lines may contribute.

## 3.4. X-rays

- Point-like sources appear: X-ray binaries, supernova remnants, and possibly weak nuclear activity.
- In M100, SN 1979C stands out as a well-known X-ray source.

The X-ray emission of M100 combines point sources (X-ray binaries and supernova remnants) with diffuse hot gas associated with dynamical processes in the arms and nucleus. Among the point sources, SN 1979C is notable, being detected decades after the explosion. RASS/XMM/Chandra show that part of the gas reaches temperatures of a few million kelvin; the nucleus may supply an additional weak component.

*Fig. 5.— Optical (DSS2 colour) image with ROSAT All-Sky Survey (RASS) X-ray contours overlaid. The position of SN 1979C is marked.*

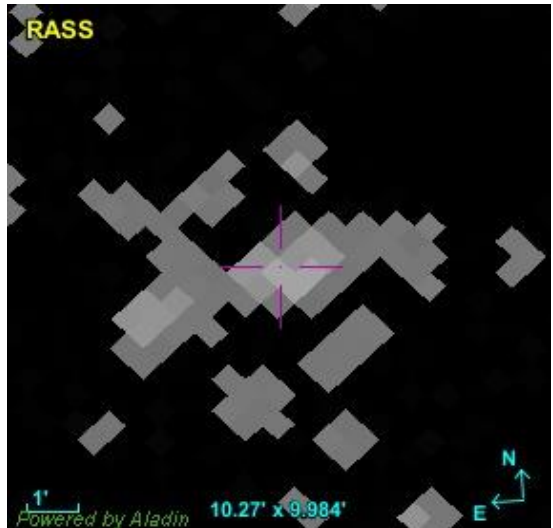

Credits. DSS2 + RASS via Aladin (CDS).

- **Optical vs. NIR:** the striking spiral pattern seen in the optical does not necessarily coincide with the **stellar-mass distribution** traced by the NIR.
- **UV vs. optical:** the UV highlights **recent star formation** that may be **obscured by dust** in the optical.
- **X-rays vs. UV:** some X-ray sources (e.g. **supernova remnants**) lie near bright UV regions, whereas others do not.
- **Take-home message:** each band tells a different part of the story; together they provide the **physical picture** of M100.

**NIR–blue correlation.** Along the arms there can be a partial correlation (blue from massive stars; NIR from heated dust plus the old stellar background). In the bulge, the correlation breaks down: the NIR is dominated by the old population, whereas the blue light is weak owing to the low current SFR. On global scales, younger systems show a tighter correlation than galaxies dominated by old populations.

## 3.5. Barred vs. unbarred spiral — comparative morphology

We now present a comparative morphological analysis of two galaxies: the barred spiral NGC 1300 and the unbarred spiral M 74. We compute elliptical isophotes, compile/estimate redshift and V-band magnitude, and then estimate the distance and infer the luminosity.

*Fig. 6.— Barred spiral galaxy NGC 1300.*

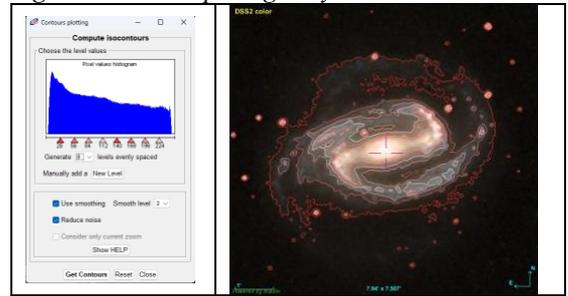

Credits. Aladin (CDS).

For **NGC 1300** we plot eight equally spaced isophotes. The central region clearly reveals a bar, and two spiral arms are seen, one more prominent than the other. The elongated bar that spans the nuclear region acts to redistribute angular momentum in the gas and stars, funnelling material toward the galactic centre. Isophotes across the bar are more flattened than those of the outer disc, confirming the presence of a bar.

*Fig. 7.— Isophotes of the spiral galaxy Messier 74.*

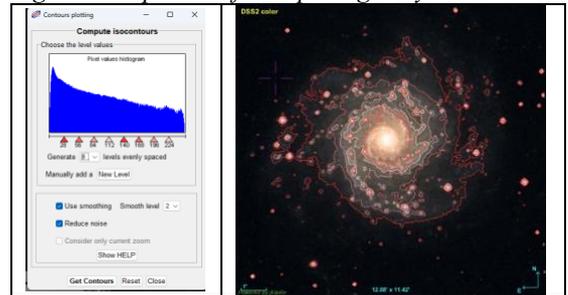

Credits. Authors' processing based on Aladin (CDS) data.

**NGC 1300 (barred spiral)**

- An elongated central structure is clearly visible in the isophotes.
- The spiral arms emerge from the ends of the bar.

- The bar signals gravitational instabilities that can influence the galaxy's secular evolution.

**M 74 (unbarred spiral)**

- The inner isophotes are more circular, with no evidence of a central bar.
- Spiral arms originate directly from the nucleus rather than from bar ends.
- This morphology suggests more distributed star formation, less driven by internal bar-induced torques.

The isophotal comparison confirms the distinct morphology of each galaxy. In terms of star formation and evolution, the central bar in NGC 1300 can act as a channel for gas inflow toward the nucleus, potentially triggering central star-formation episodes; the absence of a bar in M 74 points to a more uniform gas distribution across the disc with extended star formation along the arms.

The following table lists the radial velocity and redshift **z**—from which we can infer the distance and the relative motion—as well as the V-band apparent magnitude, which characterises the observed brightness of each galaxy.

Table 1. General properties of NGC 1300 and Messier 74

| Objeto | $V_{rad}$ | z | mv (mag) |
|---|---|---|---|
| NGC 1300 | 1570.5 km/s | 0.005252 | 10.42 |
| M 74 | 658 km/s | 0.002197 | 9.46 |

Credits. Authors' processing; values compiled from SIMBAD (CDS).

**Ellipticity and equivalent radius of M 60**

We compute the ellipticity (ε = 1−b/a) and the equivalent radius ($R_{eq} = \sqrt{ab}$) of the elliptical galaxy Messier 60 (NGC 4649), which forms a close pair and likely gravitationally interacts with the nearby companion NGC 4647.

Fig. 8.— *Isophotes of the elliptical galaxy Messier 60.*

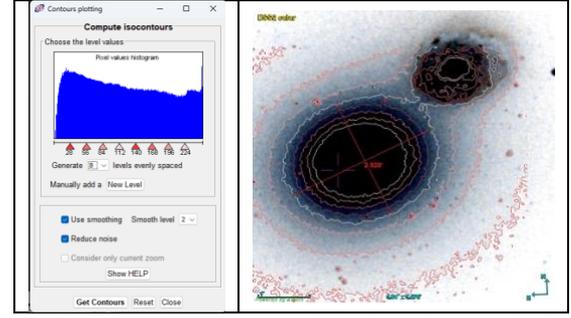

Credits. Authors' processing based on Aladin (CDS) data.

Eight equidistant intensity isophotes were identified. Of these, the third isophote, corresponding to an intensity of 64 in the cumulative histogram, was selected. On this isophote, the area on the right relative to the area on the left shows an approximate 3:1 ratio, and this was designated as $R_{25}$. To optimise the measurements, an inverted image was generated on which the major and minor axes were measured. With these values, the ellipticity and equivalent radius were computed. The results are presented below.

Table 2. Morphological characteristics of the galaxy Messier 60

| | | |
|---|---|---|
| Major axis | a | 2.928' |
| Minor axis | b | 2.369' |
| Ellipticity | $e = 1 - \dfrac{b}{a}$ | 0.1909 |
| Equivalent radius | $r = \sqrt{ab}$ | 2.6337' |

Credits. Authors' compilation from SIMBAD (CDS).

**Deriving parameters for NGC 7331**

NGC 7331 is a well-studied SA spiral galaxy *(de Vaucouleurs 1976)* located in Pegasus. Its V-band apparent magnitude is $m_V$=9.48m *(Gil de Paz et al. 2007)*.

Table 3. General properties of NGC 7331

| Object | $V_{rot}$ | z | $M_H$(mag) | $m_H$ (mag) |
|---|---|---|---|---|
| NGC 7331 | 265 km/s [1] | 0.002779 | -21.2 [2] | 6.294 |

Credits. Authors' processing based on SIMBAD (CDS) data.

---

[1] Shaun, M et al calcula que la velocidad radial de NGC 7331 es de 531 ± 10 km/s

[2] Véase Rubin, V et atl. 1964.

*Fig. 9.— Rotation curve from the H I 21 cm line.*

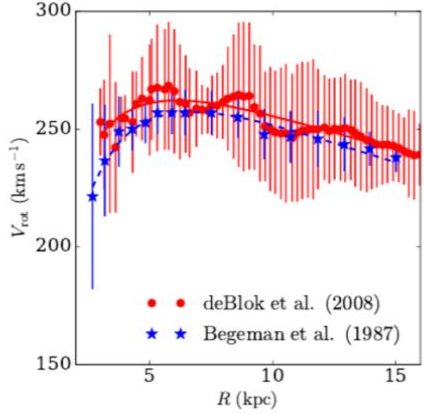

Credits. Narendra N. Patra (2018).

**Determining the distance to NGC 7331**

One way to estimate a galaxy's distance is via the Tully–Fisher (TF) relation, which links a galaxy's luminosity to its rotation velocity inferred from the width of the neutral-hydrogen 21-cm line.

**Tully–Fisher relation.**

$$L = (AV_{max})^\alpha$$

Thus, if the apparent magnitude and the rotation velocity are known, the galaxy's distance can be derived.
Substituting in the previous equation into the distance–modulus definition and solving for D gives

$$M = -23.0 - 10\log\left(\frac{V_{rot}}{200\frac{km}{s}}\right)$$

$$M = -23.0 - 10\log\left(\frac{265 km/s}{200 km/s}\right)$$

$$\log\left(\frac{265}{200}\right) = 0.1222$$

$$M = -23.0 - 10 \cdot 0.1222$$

$$M = -24.222$$

With the absolute magnitude in hand, we use the distance modulus, written as

$$\log D = \frac{m - M + 5}{5}$$

$$\log D = \frac{6.294 - (-24.222) + 5}{5}$$

$$\log D = \frac{35.516}{5}$$

$$\log D = 7.1032$$

which implies that

$$\log D = 10^{7.1032}$$

$$D = 12\,682\,357.76$$

$$\boldsymbol{D \cong 12.68\ Mpc}$$

In addition, the result can be expressed **in** light-years by using the parsec–to–light-year conversion 1 pc≈3.26 ly1:

$$D = 41\ M.a.l.$$

The distance obtained for NGC 7331, D=12.68D = 12.68D=12.68 Mpc, from the Tully–Fisher relation is consistent with previous determinations placing the galaxy at ∼12−15 Mpc (Rubin et al. 1965; HST Key Project 2000). This agreement supports the applicability of the method in our analysis.

**4.Conclusions**

**Viewing a galaxy "in layers" changes the physical interpretation.**
1. The UV–optical–NIR–X-ray comparison shows that the spectacular blue/optical arms do not necessarily trace the stellar-mass distribution (better followed in the NIR). In M100 the UV highlights recent star-forming knots, the NIR outlines the bulge/oval features, and X-ray contours reveal compact sources and hot gas (including the SN 1979C region). A multiband approach avoids conclusions based solely on short-lived luminous tracers.

2. **Isophotal morphology quantitatively discriminates bars.**
The contrast between NGC 1300 (barred) and M 74 (unbarred) shows that inner-isophote flattening and patterns identify bars and their coupling to the spiral arms**.** The bar in NGC 1300 is consistent with angular-momentum redistribution and episodes of central fueling; in M 74 the activity is more azimuthally uniform.

3. **Geometric metrics in ellipticals are stable and reproducible.**

For M 60 we measure a moderate ellipticity ε=0.19 and an equivalent radius R$_{eq}$=2.634′, consistent with an E/S0 not exactly edge-on. Repeating the measurement on neighbouring isophotes and reporting the dispersion provides an operational estimate of the uncertainty without requiring absolute calibration.

4. **Tully–Fisher works as a didactic example and exposes the bottlenecks.**
For NGC 7331, the TF exercise clarifies the workflow: W$_{obs}$≃2 v$_{rot}$, inclination correction W$_c$=W$_{obs}$/sin$i$, consistent choice of a single band (do not mix H, I, KsK_sKs), and a consistent photometric system. With the reference values used here, the distance is D∼12−13 Mpc, with the uncertainty dominated by iii, internal corrections, and the intrinsic scatter of the relation.

5. **Public-data methodology = high cost–benefit.**
A pipeline based on **open archives** (DSS2/SDSS, GALEX, 2MASS, ROSAT/XMM/Chandra), a common WCS registration, and Aladin as the working environment enabled reproducibility and homogeneous cross-band comparisons. This scheme is well suited to advanced teaching and to launching observational projects with limited resources.

6. Limitations and biases identified (and how to mitigate them)

- **PSF and resolution mismatch.** Disparate PSFs across bands can bias fine radial profiles; for quantitative radial work, match to the worst FWHM.
- **Sky background and masking.** Background subtraction and masking (foreground stars/satellites) affect outer isophotes; mitigate with robust sky estimates and explicit masks.
- **Absolute photometry.** When no calibration is available, profiles must be treated as **relative** and this should be stated explicitly.
- **Operational definitions.** Quantities such as R$_{25}$ (defined at $\mu B = 25\ mag\ arcsec^{-2}$) require a consistent band and photometric system.
- **X-rays.** RASS maps are useful for global contours, but point-source analysis requires pointed observations (XMM/Chandra).

7. **Physical implications**

- The partial decoupling between blue/UV and NIR underscores the role of dust and age gradients: the underlying stellar mass can be rounder and more bar-dominated than suggested by the optical.
- Bars emerge as engines of secular evolution, modulating the radial distribution of gas and stars and likely linked to central starbursts.
- In bright ellipticals such as M 60, smooth variations of ellipticity and PA are compatible with triaxiality and/or anisotropy gradients.

8. **Future work (direct value add)**

- **PSF** matching and photometric calibration to derive μλ (R) in physical units and compare colour gradients.
- **Bulge–disc–bar decomposition** (e.g. GALFIT/IMFIT) and stellar-mass maps via colour-dependent M/LM/LM/L.
- **SFR estimates** with GALEX plus attenuation corrections (e.g. Calzetti curves) and comparison with NIR tracers.
- **Bar strength** QbQ_bQb from NIR potential maps and torques; connection to ring/spiral morphology.
- Integrate **H I/CO kinematics** (line widths, velocity fields) to strengthen Tully–Fisher and probe asymmetries.
- Replace RASS with XMM/Chandra pointings where available to build source catalogues and thermal maps.
- **Broaden the sample** across environments (field vs. clusters) to isolate tidal effects and environmental coupling.

In summary, a multiband reading mitigates biases: UV vs. NIR separates young tracers from the stellar mass (Gil de Paz et al. 2007; Elmegreen et al. 2004); X-ray contours add the energetic component (Palumbo et al. 1981; Kaaret 2001). Isophotes discriminate bars and their secular role (Knapen et al. 1999). Geometric metrics in ellipticals are stable with archival data (Tonry & Schneider 1988). The Tully–Fisher distance for NGC 7331 is consistent with previous scales (Rubin et al. 1965; HST Key Project 2000) and with recent rotation-curve work (Patra 2018).